# A Stability Analysis for the Acceleration-based Robust Position Control of Robot Manipulators via Disturbance Observer

Emre Sariyildiz, *Member, IEEE*, Hiromu Sekiguchi, Takahiro Nozaki, *Member, IEEE*, Barkan Ugurlu, *Member, IEEE*, Kouhei Ohnishi, *Life Fellow, IEEE*

*Abstract* — This paper proposes a new nonlinear stability analysis for the acceleration-based robust position control of robot manipulators by using Disturbance Observer (DOb). It is shown that if the nominal inertia matrix is properly tuned in the design of DOb, then the position error asymptotically goes to zero in regulation control and is uniformly ultimately bounded in trajectory tracking control. As the bandwidth of DOb and the nominal inertia matrix are increased, the bound of error shrinks, i.e., the robust stability and performance of the position control system are improved. However, neither the bandwidth of DOb nor the nominal inertia matrix can be freely increased due to practical design constraints, e.g., the robust position controller becomes more noise sensitive when they are increased. The proposed stability analysis provides insights regarding the dynamic behavior of DOb-based robust motion control systems. It is theoretically and experimentally proved that non-diagonal elements of the nominal inertia matrix are useful to improve the stability and adjust the trade-off between the robustness and noise sensitivity. The validity of the proposal is verified by simulation and experimental results.

## I. Introduction

ROBUSTNESS is crucial to achieve high performance and accuracy in the motion control applications of mechatronic systems, such as high-resolution CNC machines, precise assembly robots, high-speed hard disk drives and magnetic levitation [1]–[6]. While the robustness of a motion control system can be simply improved by increasing controller gains, there are many practical issues in high gain feedback control. For example, the unmodeled high frequency dynamics of a mechanical system may be excited, the mechanical resonance may destabilize the system and noise and energy consumption are increased. Model-based controllers, such as feedback linearization, are generally used to eliminate plant nonlinearities [7]. However, it is not an easy task to derive the exact dynamic models of many robotic and mechatronic systems. In addition to plant nonlinearities, the performance of a motion control system may suffer from unknown disturbances such as external loads. Therefore, a robust motion controller is required to suppress internal and external disturbances in practice.

To tackle internal and external disturbances of motion control systems, several robust motion controllers (e.g., H∞ and sliding mode controllers) have been proposed in the literature [8]–[12]. Among them, acceleration-based control (ABC) is one of the most popular robust motion controller synthesis techniques due to its simplicity and efficacy [8, 9, 13]. In ABC, the robustness of the motion controller is intuitively achieved by attenuating disturbances with their estimations via a DOb in an inner-loop. The performance of the motion controller is tuned by considering only the nominal plant dynamics in an outer-loop as the plant-model mismatches and external disturbances are well-suppressed in the inner-loop. An ABC system has two-degrees-of-freedom; i.e., its robustness and performance can be independently adjusted by using DOb and performance controller in the inner and outer loops, respectively.

Although many successful motion control applications of DOb have been performed in the last three decades, there is still a lack of sufficient analysis and synthesis techniques for DOb-based robust control [13]–[18]. Therefore, its stability and performance generally depend on designers' experiences. A DOb-based robust motion control system is conventionally analyzed by using linear control methods [8, 9, 16, 17]. To eliminate the interactive disturbances of robot dynamics, a robust decentralized motion controller is synthesized by using a diagonal nominal inertia matrix in the design of DOb [16]. The asymptotic stability of the decentralized ABC system is proved for each joint by using linear stability analysis methods such as root-locus [9, 16]. However, it does not reflect the real dynamic behavior of the robust motion control system as many robot manipulators have highly nonlinear and coupled dynamics. To improve the stability and performance of the DOb-based robust position control of robot manipulators, more realistic analysis and synthesis techniques should be proposed [18]–[20]. A nonlinear DOb synthesis was proposed to improve the performance of disturbance estimation by using the known nonlinear dynamics of robots in the design of DOb [20]. More advanced robust control analysis and synthesis techniques since then have been proposed for the observer-based robust control systems [21, 22]. A comprehensive survey on DOb-based robust control can be found in [18]. The advanced techniques generally neglect the practical limitations of DOb-based robust motion control systems. Recently, the authors proposed a linear analysis and synthesis technique for a DOb-based robust motion control system by considering its practical design constraints such as noise sensitivity [9]. However, the proposed practical method neglects the nonlinear dynamics of robot manipulators.

This paper proposes a new nonlinear stability analysis for the ABC of robot manipulators by using DOb. It theoretically and experimentally verifies that not only the bandwidth of DOb but also the nominal inertia matrix significantly influences the stability and performance of the robust motion control system. To prove the stability, the authors first show the equivalence of the ABC and passivity-based position control systems; and a passivity-based stability analysis is proposed in *Theorem 1* [19]. Since it is not an easy task to show the passivity of the ABC of a robot manipulator, *Theorem 1* is generally not applicable. A more practical stability analysis is proposed by relaxing the strict passivity condition in *Theorem 2* and *Theorem 3*. It is shown that the ABC system becomes unstable if the nominal inertia matrix is smaller than the uncertain inertia matrix and the bandwidth of DOb cannot compensate the inertia variations. If the nominal inertia matrix is larger than the uncertain inertia matrix, then the position error is uniformly ultimately bounded in trajectory tracking control (*Theorem 2*) and asymptotically goes to zero in regulation control (*Theorem 3*). The bound of the position error shrinks as the bandwidth of DOb and/or the nominal inertia matrix are increased. Although the robust stability and performance can be directly improved by increasing either the bandwidth of DOb or the nominal inertia matrix in theory, they are limited by practical design constraints such as sampling time in real implementations. As the bandwidth of DOb and/or nominal inertia matrix are increased, the robust position controller becomes more noise sensitive. Therefore, the trade-off between the robustness and noise sensitivity should be considered in the design of the ABC systems. This paper shows that not only the diagonal but also the non-diagonal terms of the nominal inertia matrix can be used to improve the stability of the ABC systems. The latter are also useful to adjust the trade-off between the robustness and noise sensitivity. The validity of the proposal is verified by simulations and experiments.

The theoretical framework of this paper was first presented in [23]. However, it lacks experimental verification and discussion on tuning the robust position controller. This paper extends the previous theoretical work by providing new design tools for DOb; e.g., the non-diagonal terms of the nominal inertia matrix are first considered to tune the robustness and stability without degrading the noise sensitivity. Moreover, theoretical results are first experimentally verified by using a redundant planar robot arm in this paper.

The rest of the paper is organized as follows. In Section II, preliminaries are given for the dynamic model of robot manipulators. They are used to prove the stability of the robust position control system. In Section III, DOb and the acceleration-based robust position control system are briefly explained. In Section IV, a new nonlinear stability analysis is proposed for ABC systems. In Section V, simulation and experimental results are presented. The paper is concluded in Section VI.

## II. PRELIMINARIES

The dynamic model of a robot manipulator with n-degrees-of-freedom is derived by using Euler-Lagrange formulation and expressed in joint space as follows:

$$\mathbf{M}(\mathbf{q})\ddot{\mathbf{q}} + \mathbf{C}(\mathbf{q},\dot{\mathbf{q}})\dot{\mathbf{q}} + \mathbf{g}(\mathbf{q}) = \boldsymbol{\tau} - \boldsymbol{\tau}^{\text{frc}} - \boldsymbol{\tau}^{\text{load}} \quad (1)$$

where $\mathbf{M}(\mathbf{q}) \in \mathbb{R}^{n \times n}$ is the inertia matrix; $\mathbf{C}(\mathbf{q},\dot{\mathbf{q}})\dot{\mathbf{q}} \in \mathbb{R}^n$ is the vector of Coriolis and centrifugal torques; $\mathbf{g}(\mathbf{q}) \in \mathbb{R}^n$ is the vector of the gravitational torques that is obtained as the gradient of the robot's potential energy due to gravity; $\boldsymbol{\tau} \in \mathbb{R}^n$ is the vector of generalized torques in joint space; $\boldsymbol{\tau}^{\text{frc}}$ and $\boldsymbol{\tau}^{\text{load}} \in \mathbb{R}^n$ are respectively the vectors of friction and load torques; and $\mathbf{q}, \dot{\mathbf{q}}$ and $\ddot{\mathbf{q}} \in \mathbb{R}^n$ are the vectors of angle, velocity and acceleration of the robot manipulator in joint space, respectively.

Let us assume that the robot manipulator under consideration has only revolute joints and reference trajectories, $\mathbf{q}^{\text{ref}}(t), \dot{\mathbf{q}}^{\text{ref}}(t)$ and $\ddot{\mathbf{q}}^{\text{ref}} \in \mathbb{R}^n$, are continuous and bounded. Let us also assume that the friction and load torques are bounded by $\|\boldsymbol{\tau}^{\text{frc}}\|_2 \leq \beta_{\text{frc}}^{\max}$ and $\|\boldsymbol{\tau}^{\text{load}}\|_2 \leq \beta_{\text{load}}^{\max}$ where $\beta_{\text{frc}}^{\max}$ and $\beta_{\text{load}}^{\max}$ are positive real constants, and $\|\bullet\|_2$ is Euclidean norm of vector $\bullet \in \mathbb{R}^n$.

Eq. (1) has the following important properties that will be used in the analysis and synthesis of the robust motion controller [24].

*Property1:* $\mathbf{M}(\mathbf{q})$ is a positive definite and symmetric inertia matrix that satisfies

$$\beta_{\mathbf{M}}^{\min} \mathbf{I} \leq \underline{\sigma}(\mathbf{M}(\mathbf{q})) \mathbf{I} \leq \mathbf{M}(\mathbf{q}) \leq \bar{\sigma}(\mathbf{M}(\mathbf{q})) \mathbf{I} \leq \beta_{\mathbf{M}}^{\max} \mathbf{I} \quad (2)$$

where $\underline{\sigma}(\bullet)$ and $\bar{\sigma}(\bullet)$ are minimum and maximum eigenvalues of $\bullet$, respectively; $\beta_{\mathbf{M}}^{\min}$ and $\beta_{\mathbf{M}}^{\max}$ are positive real constants; and $\mathbf{I} \in \mathbb{R}^{n \times n}$ is an identity matrix.

*Property2:* $\|\mathbf{C}(\mathbf{q},\dot{\mathbf{q}})\dot{\mathbf{q}}^*\|_2 \leq \beta_{\mathbf{C}} \|\dot{\mathbf{q}}\| \|\dot{\mathbf{q}}^*\|$ where $\beta_{\mathbf{C}}$ is a positive real constant; $\dot{\mathbf{q}}^* \in \mathbb{R}^n$ is a continuous and bounded vector.

*Property3:* $\|\mathbf{g}(\mathbf{q})\|_2 \leq \beta_{\mathbf{g}}$ where $\beta_{\mathbf{g}}$ is a positive real constant.

*Property4:* $\mathbf{x}^T \left( \frac{d}{dt}(\mathbf{M}(\mathbf{q})) - 2\mathbf{C}(\mathbf{q},\dot{\mathbf{q}}) \right) \mathbf{x} = 0$ where $\mathbf{x} \in \mathbb{R}^n$; and $\frac{d}{dt}(\mathbf{M}(\mathbf{q})) - 2\mathbf{C}(\mathbf{q},\dot{\mathbf{q}})$ is a skew symmetric matrix.

## III. ROBUST POSITION CONTROL VIA DOB

In this section, the acceleration-based robust position controller, which cancels/suppresses disturbances via DOb, is briefly explained.

### A. Disturbance Observer

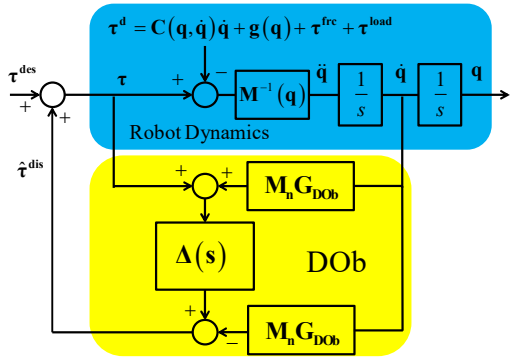

Fig. 1. Block diagram of a DOb-based robust motion control system.

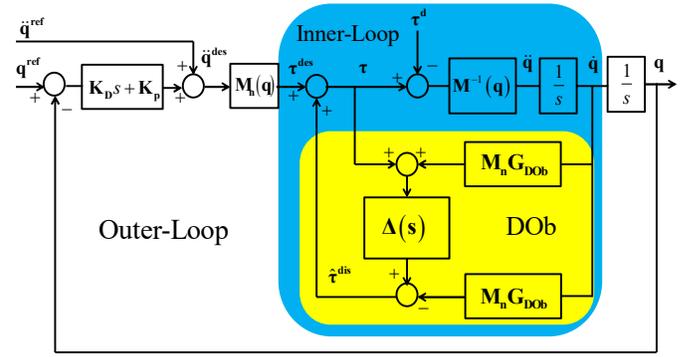

Fig. 2. Block diagram of an acceleration-based robust position control system.

A block diagram for a DOb-based motion control system is illustrated in Fig. 1. In this figure, $\mathbf{M_n} \in \mathbb{R}^{n \times n}$ represents the nominal inertia matrix which holds $\beta_{\mathbf{M_n}}^{min}\mathbf{I} \leq \mathbf{M_n} \leq \beta_{\mathbf{M_n}}^{max}\mathbf{I}$ where $\beta_{\mathbf{M_n}}^{min}$ and $\beta_{\mathbf{M_n}}^{max}$ are real positive numbers; $\boldsymbol{\tau}^{des} \in \mathbb{R}^n$ represents the desired generalized torques in joint space; $\mathbf{G_{DOb}} = Diag\left\{\left[g_{DOb_1}, g_{DOb_2}, \cdots, g_{DOb_n}\right]^T\right\} \in \mathbb{R}^{n \times n}$ where $Diag\{\bullet\}: \mathbb{R}^n \to \mathbb{R}^{n \times n}$ is a diagonal matrix of vector $\bullet \in \mathbb{R}^n$, and $g_{DOb_i} \in \mathbb{R}$ represents the bandwidth of DOb for the $i^{th}$ joint of the robot; $\boldsymbol{\Delta}(\mathbf{s}) = Diag\left\{\left[\frac{g_{DOb_1}}{s+g_{DOb_1}}, \frac{g_{DOb_2}}{s+g_{DOb_2}}, \cdots, \frac{g_{DOb_n}}{s+g_{DOb_n}}\right]\right\} \in \mathbb{R}^{n \times n}$ represents the matrix of the low-pass filter of DOb in which $s$ represents complex Laplace variable; $\boldsymbol{\tau}^d \in \mathbb{R}^n$ represents the disturbance vector which includes disturbances due to the external load, friction and nonlinear robot dynamics which is excluded from the nominal dynamic model in the design of DOb; and $\boldsymbol{\tau}^{dis} \in \mathbb{R}^n$ represents the disturbance vector which includes $\boldsymbol{\tau}^d$ and the disturbances due to the nominal and exact inertia's mismatch, i.e., $\boldsymbol{\tau}^{dis} = (\mathbf{M_n} - \mathbf{M(q)})\ddot{\mathbf{q}} + \boldsymbol{\tau}^d$.

The estimation of the disturbance vector $\boldsymbol{\tau}^{dis}$ includes not only external but also internal disturbances and is expressed by using

$$\hat{\boldsymbol{\tau}}^{dis} = \boldsymbol{\Delta}(\mathbf{s})\boldsymbol{\tau}^{dis} \qquad (3)$$

where $\hat{\boldsymbol{\tau}}^{dis} \in \mathbb{R}^n$ represents the estimation of $\boldsymbol{\tau}^{dis}$ [7, 8].

The dynamic equations of a DOb-based robust motion control system are directly derived from Fig. 1 as follows:

$$\hat{\boldsymbol{\tau}}^{dis} = \boldsymbol{\Delta}(\mathbf{s})\left(\hat{\boldsymbol{\tau}}^{dis} + \boldsymbol{\tau}^{des} - \mathbf{M_n}\ddot{\mathbf{q}}\right) \qquad (4)$$

$$\hat{\boldsymbol{\tau}}^{dis} = \hat{\boldsymbol{\Delta}}(\mathbf{s})\left(\boldsymbol{\tau}^{des} - \mathbf{M_n}\ddot{\mathbf{q}}\right) \qquad (5)$$

where $\hat{\boldsymbol{\Delta}}(\mathbf{s}) = Diag\left\{\left[\frac{g_{DOb_1}}{s}, \frac{g_{DOb_2}}{s}, \cdots, \frac{g_{DOb_n}}{s}\right]\right\} \in \mathbb{R}^{n \times n}$.

### B. Acceleration-based Robust Position Controller

A block diagram for an acceleration-based robust position control system is illustrated in Fig. 2. In this figure, $\mathbf{K_D}$ and $\mathbf{K_P} \in \mathbb{R}^{n \times n}$ represent the diagonal velocity and position gain matrices of the PD controller, respectively; $\ddot{\mathbf{q}}^{des} \in \mathbb{R}^n$ represents the desired acceleration vector; and $\mathbf{q}^{ref}, \dot{\mathbf{q}}^{ref}$ and $\ddot{\mathbf{q}}^{ref} \in \mathbb{R}^n$ represent the reference vectors of angle, velocity and acceleration, respectively. The other parameters are same as defined earlier. In ABC, all disturbances are cancelled via DOb in the inner-loop so that the joints of a robot manipulator can precisely track the desired acceleration vector, i.e., $\ddot{\mathbf{q}} = \ddot{\mathbf{q}}^{des}$ as $t \to \infty$ [8, 9].

There are four control parameters, $\boldsymbol{\Delta}(\mathbf{s})$, $\mathbf{M_n}$, $\mathbf{K_D}$ and $\mathbf{K_P}$, that should be tuned in the design of the acceleration-based robust position control systems.

The design of the low-pass filter of DOb $(\boldsymbol{\Delta}(\mathbf{s}))$ has been widely studied in the literature. It is a well-known fact that the higher the bandwidth of the low-pass filter/DOb is, the more the robustness improves. However, its bandwidth is limited by some practical constraints such as noise and sampling time [9, 13]. Therefore, the bandwidth of DOb is set as high as possible in practical applications.

Unlike the bandwidth of DOb, tuning the parameters of the nominal inertia matrix $(\mathbf{M_n})$ is not straightforward in the design of the acceleration-based robust position control systems. Conventionally, a diagonal nominal inertia matrix is used so that decentralized robust position controllers are independently designed at each joint [9, 16]. The analysis and synthesis of the decentralized robust motion controllers are generally conducted by using linear control methods due to simplicity. However, the linear control methods are not useful in the ABC of robot manipulators as multi-degrees-of-freedom robots have highly nonlinear and coupled dynamic models [17]. Furthermore, the non-diagonal terms of the nominal inertia matrix, which can be used to adjust the trade-off between the robustness and noise sensitivity, are generally neglected in the conventional design of DOb.

The parameters of the outer-loop controller, i.e., $\mathbf{K_D}$ and $\mathbf{K_P}$, are tuned by only considering the nominal system dynamics as external disturbances and plant uncertainties are well-attenuated by DOb in the inner-loop.

### IV. STABILITY ANALYSIS OF THE ABC SYSTEM

In this section, a new nonlinear stability analysis is proposed for the ABC of robot manipulators.

Let us assume that the low-pass filters of DOb are designed by using $g_{DOb_1} = g_{DOb_2} = \cdots = g_{DOb_n} = g_{DOb}$. If the control signal is generated by using the acceleration-based robust position controller that is explained in Section III, then the dynamic model of the robot manipulator is derived as follows:

$$\mathbf{M}(\mathbf{q})\ddot{\mathbf{q}} + \mathbf{C}(\mathbf{q},\dot{\mathbf{q}})\dot{\mathbf{q}} + \mathbf{g}(\mathbf{q}) = \boldsymbol{\tau}^{des} + \hat{\boldsymbol{\tau}}^{dis} - \boldsymbol{\tau}^{frc} - \boldsymbol{\tau}^{load} \quad (6)$$

The vector of the desired joint torques is

$$\boldsymbol{\tau}^{des} = \mathbf{M}_\mathbf{n}\ddot{\mathbf{q}}^{des} \quad (7)$$

where $\ddot{\mathbf{q}}^{des} = \ddot{\mathbf{q}}^{ref} - \mathbf{K}_\mathbf{D}(\dot{\mathbf{q}} - \dot{\mathbf{q}}^{ref}) - \mathbf{K}_\mathbf{P}(\mathbf{q} - \mathbf{q}^{ref}) = \ddot{\mathbf{q}}^{ref} - \mathbf{K}_\mathbf{D}\dot{\mathbf{e}} - \mathbf{K}_\mathbf{P}\mathbf{e}$.

The vector of the disturbance estimation is derived by substituting Eq. (7) into Eq. (5) as follows:

$$\hat{\boldsymbol{\tau}}^{dis} = g_{DOb}\mathbf{M}_\mathbf{n}(\dot{\mathbf{q}}^{des} - \dot{\mathbf{q}}) \quad (8)$$

If Eq. (7) and Eq. (8) are applied into Eq. (6), then

$$\mathbf{M}(\mathbf{q})\ddot{\mathbf{q}} + \mathbf{C}(\mathbf{q},\dot{\mathbf{q}})\dot{\mathbf{q}} + \mathbf{g}(\mathbf{q}) = \mathbf{M}_\mathbf{n}\ddot{\mathbf{q}}^{des} + g_{DOb}\mathbf{M}_\mathbf{n}(\dot{\mathbf{q}}^{des} - \dot{\mathbf{q}}) - \boldsymbol{\tau}^{frc} - \boldsymbol{\tau}^{load} \quad (9)$$

Let us define a dynamic error vector by using $\mathbf{e}_\mathbf{D} = \dot{\mathbf{q}} - \dot{\mathbf{q}}^{des}$. Eq. (9) can be rewritten in terms of the dynamic error vector as follows:

$$\mathbf{M}(\mathbf{q})\dot{\mathbf{e}}_\mathbf{D} + \mathbf{C}(\mathbf{q},\dot{\mathbf{q}})\mathbf{e}_\mathbf{D} + g_{DOb}\mathbf{M}_\mathbf{n}\mathbf{e}_\mathbf{D} = -\boldsymbol{\psi} \quad (10)$$

where $\boldsymbol{\psi} = \Delta\mathbf{M}(\mathbf{q})\ddot{\mathbf{q}}^{des} + \mathbf{C}(\mathbf{q},\dot{\mathbf{q}})\dot{\mathbf{q}}^{des} + \mathbf{g}(\mathbf{q}) + \boldsymbol{\tau}^{frc} + \boldsymbol{\tau}^{load}$ represents the disturbance vector; $\mathbf{e}_\mathbf{D} = \dot{\mathbf{e}} + \mathbf{K}_\mathbf{D}\mathbf{e} + \mathbf{K}_\mathbf{P}\int\mathbf{e}dt$ represents the dynamic error vector; $\mathbf{e} = \mathbf{q} - \mathbf{q}^{ref}$ represents the position error vector; and $\Delta\mathbf{M}(\mathbf{q}) = \mathbf{M}(\mathbf{q}) - \mathbf{M}_\mathbf{n}$ is a bounded matrix which represents the difference between uncertain and nominal inertiae.

Eq. (10) shows that the passivity and acceleration based position control systems have same error dynamics [19]. Therefore, the following theorem can be directly applied to the acceleration-based robust position control systems.

**Theorem 1:** If the mapping $-\mathbf{e}_\mathbf{D} \to -\boldsymbol{\psi}$ is passive, i.e.,

$$\int_0^{t_f} \mathbf{e}_\mathbf{D}^T(t)\boldsymbol{\psi}(t)dt \geq -\phi \quad (11)$$

for all $t_f$ and for some $\phi \geq 0$., then $\mathbf{e}_\mathbf{D} \in L_2^n \cap L_\infty^n, \dot{\mathbf{e}}_\mathbf{D} \in L_2^n$, $\mathbf{e}_\mathbf{D}$ is continuous and $\mathbf{e}_\mathbf{D} \to 0$ as $t \to \infty$.

**Proof:** The inverse of the transfer function between $\mathbf{e}_\mathbf{D}$ and $\mathbf{e}$, i.e.,

$$\mathbf{e}_\mathbf{D} = s + \mathbf{K}_\mathbf{D} + \mathbf{K}_\mathbf{P}\frac{1}{s}\mathbf{e} \quad (12)$$

is stable and strictly proper. Therefore, the acceleration-based robust position control system is asymptotically stable if the mapping $-\mathbf{e}_\mathbf{D} \to -\boldsymbol{\psi}$ is passive [19]. Q.E.D.

**Theorem 1** provides us a basic insight into the stability of the acceleration-based robust position control systems. For example, the asymptotic stability is achieved if $\boldsymbol{\psi}=0$. However, this assumption is impractical as many robotic systems suffer from different disturbances, such as friction, load and modelling uncertainties. Moreover, gravity, Coriolis and centrifugal forces cannot be ignored in many applications. In general, the passivity mapping cannot be easily proved due to the complex integration in Eq. (11).

To show the stability of the ABC for a general class of robotic systems, the main theorem of the paper is proposed as follows:

**Theorem 2:** The origin of the system that is given in Eq. (10) is uniformly ultimately bounded with respect to the set $B_\Gamma = \{\|\mathbf{e}_\mathbf{D}\|_2 > \Gamma\}$ where

$$\Gamma = \frac{\beta_{\Delta\mathbf{M}}^{max}\|\ddot{\mathbf{q}}^{des}\|_2 + \beta_C\|\dot{\mathbf{q}}\|_2\|\dot{\mathbf{q}}^{des}\|_2 + \beta_g + \beta_{frc}^{max} + \beta_{load}^{max}}{g_{DOb}\beta_{\mathbf{M}_\mathbf{n}}^{min}} \quad (13)$$

if DOb is designed by using $\mathbf{M}_\mathbf{n} \geq \mathbf{M}(\mathbf{q})$, i.e., $\Delta\mathbf{M}(\mathbf{q}) \leq 0$. In Eq. (13), $\beta_{\Delta\mathbf{M}}^{max}$ represents the upper bound of $\Delta\mathbf{M}(\mathbf{q})$.

**Proof:** Let us consider the Lyapunov function candidate by using

$$V = \frac{1}{2}\mathbf{e}_\mathbf{D}^T\mathbf{M}(\mathbf{q})\mathbf{e}_\mathbf{D} \quad (14)$$

The time derivative of Eq. (14) is derived as follows:

$$\dot{V} = \mathbf{e}_\mathbf{D}^T\mathbf{M}(\mathbf{q})\dot{\mathbf{e}}_\mathbf{D} + \frac{1}{2}\mathbf{e}_\mathbf{D}^T\dot{\mathbf{M}}(\mathbf{q})\mathbf{e}_\mathbf{D} \quad (15)$$

$$\dot{V} = -\mathbf{e}_\mathbf{D}^T\boldsymbol{\psi} - g_{DOb}\mathbf{e}_\mathbf{D}^T\mathbf{M}_\mathbf{n}\mathbf{e}_\mathbf{D} + \frac{1}{2}\mathbf{e}_\mathbf{D}^T(\dot{\mathbf{M}}(\mathbf{q}) - 2\mathbf{C}(\mathbf{q},\dot{\mathbf{q}}))\mathbf{e}_\mathbf{D} \quad (16)$$

If *Property 4* is applied into Eq. (16), then

$$\dot{V} = -g_{DOb}\mathbf{e}_\mathbf{D}^T\mathbf{M}_\mathbf{n}\mathbf{e}_\mathbf{D} - \mathbf{e}_\mathbf{D}^T\boldsymbol{\psi} = -g_{DOb}\mathbf{e}_\mathbf{D}^T\mathbf{M}_\mathbf{n}\mathbf{e}_\mathbf{D} - \mathbf{e}_\mathbf{D}^T\Delta\mathbf{M}(\mathbf{q})\ddot{\mathbf{q}}^{des} - \mathbf{e}_\mathbf{D}^T\{\mathbf{C}(\mathbf{q},\dot{\mathbf{q}})\dot{\mathbf{q}}^{des} + \mathbf{g}(\mathbf{q}) + \boldsymbol{\tau}^{frc} + \boldsymbol{\tau}^{load}\} \quad (17)$$

The time derivative of the Lyapunov function is smaller than zero if the following inequality holds.

$$g_{DOb}\mathbf{e}_\mathbf{D}^T\mathbf{M}_\mathbf{n}\mathbf{e}_\mathbf{D} \geq -\mathbf{e}_\mathbf{D}^T(\Delta\mathbf{M}(\mathbf{q})\ddot{\mathbf{q}}^{des} + \mathbf{C}(\mathbf{q},\dot{\mathbf{q}})\dot{\mathbf{q}}^{des} + \mathbf{g}(\mathbf{q}) + \boldsymbol{\tau}^{frc} + \boldsymbol{\tau}^{load}) \quad (18)$$

The sufficient condition of the stability is derived by using *Property 1, Property 2 and Property 3* as follows:

$$\dot{V} \leq$$
$$(-g_{DOb}\beta_{\mathbf{M}_\mathbf{n}}^{min}\|\mathbf{e}_\mathbf{D}\|_2 + \beta_{\Delta\mathbf{M}}^{max}\|\ddot{\mathbf{q}}^{des}\|_2 + \beta_C\|\dot{\mathbf{q}}\|_2\|\dot{\mathbf{q}}^{des}\|_2 + \beta_g + \beta_{frc}^{max} + \beta_{load}^{max})\|\mathbf{e}_\mathbf{D}\|_2 \quad (19)$$
$$\leq 0$$

Eq. (19) shows that the time derivative of the Lyapunov function is negative outside of the compact set $B_\Gamma = \{\|\mathbf{e}_\mathbf{D}\|_2 > \Gamma\}$ where $\Gamma$ is given in Eq. (13). Therefore, all solutions with initial point outside of $B_\Gamma$ enter this set within a finite time, and remain inside the set for future time. In other words, the dynamic error is uniformly ultimately bounded with respect to the set $B_\Gamma$. Q.E.D.

**Remark 1:** Eq. (19) shows that as the bandwidth of DOb and/or nominal inertia matrix are increased, the Lyapunov function decreases faster; i.e., the stability of the robust position control system is improved.

**Remark 2:** The radius of the compact set $B_\Gamma$ can be directly adjusted by tuning the bandwidth of DOb and/or nominal inertia matrix. Eq. (13) shows that as the bandwidth of DOb and/or nominal inertia matrix are increased, the radius of the compact set $B_\Gamma$ shrinks, and the error gets smaller.

***Remark 3:*** The stability of the acceleration-based robust position control system is improved by using $\mathbf{M_n} \geq \mathbf{M(q)}$, i.e., $\Delta \mathbf{M(q)} \leq \mathbf{0}$.

Let us consider the first two terms of the right-hand side of Eq. (17). The first one shows that the stability of the robust position control system is improved by increasing $\mathbf{M_n}$ and the bandwidth of DOb. However, how the stability is influenced by the second term is not as clear as the first one.

If it is assumed that $\mathbf{M_n(q)} = \mathbf{M(q)}$, then the second term is cancelled out and the stability is not influenced by the fluctuations of the desired acceleration. However, this assumption is generally impractical as deriving the exact dynamic models of many robots' inertia matrices is not an easy task due to their highly nonlinear and complex dynamic structures. In practice, the stability is influenced by $\mathbf{e_D}^T \Delta \mathbf{M(q)} \ddot{\mathbf{q}}^{des}$ in Eq. (17).

If *Property 1* is applied to $\Delta \mathbf{M(q)}$, then

$$\beta_{\Delta M}^{min} \mathbf{I} \leq \Delta \mathbf{M(q)} \leq \beta_{\Delta M}^{max} \mathbf{I} \quad \text{when} \quad \Delta \mathbf{M(q)} > \mathbf{0} \qquad (20)$$

$$-\beta_{\Delta M}^{max} \mathbf{I} \leq \Delta \mathbf{M(q)} \leq -\beta_{\Delta M}^{min} \mathbf{I} \quad \text{when} \quad \Delta \mathbf{M(q)} < \mathbf{0} \qquad (21)$$

where $\beta_{\Delta M}^{min}$ and $\beta_{\Delta M}^{max}$ are real positive numbers, and $\beta_{\Delta M}^{max} \geq \beta_{\Delta M}^{min}$.

Since the reference trajectory is continuous and bounded, the dynamic error $\left(\mathbf{e_D} = \dot{\mathbf{e}} + \mathbf{K_D} \mathbf{e} + \mathbf{K_P} \int \mathbf{e} dt\right)$ and the desired acceleration $\left(\ddot{\mathbf{q}}^{des} = \ddot{\mathbf{q}}^{ref} - \mathbf{K_D} \dot{\mathbf{e}} - \mathbf{K_P} \mathbf{e}\right)$ are also continuous and bounded when the robust position control system is stable. As the position control error $\left(\mathbf{e} = \mathbf{q} - \mathbf{q}^{ref}\right)$ increases or decreases, $\mathbf{e_D}$ and $\ddot{\mathbf{q}}^{des}$ tends to have different signs, i.e., $\mathbf{e_D}^T \ddot{\mathbf{q}}^{des} < 0$. For example, if $\mathbf{e} > 0$ and $\dot{\mathbf{e}} > 0$, then $\ddot{\mathbf{q}}^{des}$ keeps decreasing yet $\mathbf{e_D}$ keeps increasing; however, if $\mathbf{e} < 0$ and $\dot{\mathbf{e}} < 0$, then $\ddot{\mathbf{q}}^{des}$ keeps increasing yet $\mathbf{e_D}$ keeps decreasing. Therefore, the second term of the right-hand side of Eq. (17) holds the following inequalities when the position control error gets larger.

$$-\beta_{\Delta M}^{max} \left\| \mathbf{e_D}^T \ddot{\mathbf{q}}^{des} \right\| \leq \mathbf{e_D}^T \Delta \mathbf{M(q)} \ddot{\mathbf{q}}^{des} \leq -\beta_{\Delta M}^{min} \left\| \mathbf{e_D}^T \ddot{\mathbf{q}}^{des} \right\| \qquad (22)$$

when $\Delta \mathbf{M(q)} > \mathbf{0}$, and

$$\beta_{\Delta M}^{min} \left\| \mathbf{e_D}^T \ddot{\mathbf{q}}^{des} \right\| \leq \mathbf{e_D}^T \Delta \mathbf{M(q)} \ddot{\mathbf{q}}^{des} \leq \beta_{\Delta M}^{max} \left\| \mathbf{e_D}^T \ddot{\mathbf{q}}^{des} \right\| \qquad (23)$$

when $\Delta \mathbf{M(q)} < \mathbf{0}$.

By substituting Eq. (22) and Eq. (23) into Eq. (17), it can be easily shown that the stability of the acceleration-based robust position control systems is improved by using $\mathbf{M_n} \geq \mathbf{M(q)}$, i.e., $\Delta \mathbf{M(q)} \leq \mathbf{0}$.

***Remark 4:*** Eq. (13) shows that not only the robustness but also the stability of the acceleration-based position control systems is improved by increasing the bandwidth of DOb.

***Remark 5:*** Neither the bandwidth of DOb nor the nominal inertia matrix can be freely increased in practice. For example, the upper bounds of the bandwidth of DOb and nominal inertia can be analytically derived for servo systems by using

$$2 \frac{M_n}{M} g_{DOb} \leq g_v \qquad (24)$$

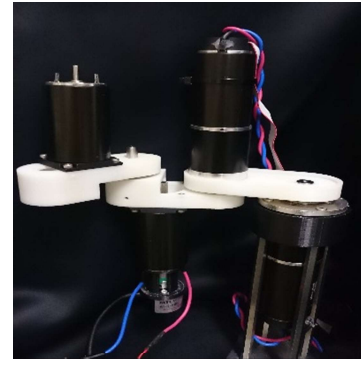

Fig. 3. Redundant planar robot manipulator.

TABLE I
SPECIFICATIONS OF EXPERIMENTAL SETUP

| Parameter | VALUE | Description |
|---|---|---|
| $l_1, l_2, l_3$ | 0.06, 0.06, 0.06 (m) | Lengths of the links |
| $m_1, m_2, m_3$ | 0.67, 0.67, 0.62 (kg) | Masses of the links |
| $I_1, I_2, I_3$ | 624, 624, 622 (gcm$^2$) | Inertiae of the links |

where $M$ and $M_n \in \mathbb{R}$ represent the uncertain and nominal inertiae, respectively; and $g_v$ represents the bandwidth of the velocity measurement [9].

Let us now consider the regulation control, i.e., point to point control, problem of robot manipulators. The reference vectors are generated by using $\ddot{\mathbf{q}}^{ref}, \dot{\mathbf{q}}^{ref} \to \mathbf{0}$, $\mathbf{q}^{ref} \to \mathbf{q}^{fixed}$ and $\dot{\mathbf{q}}^{fixed} = \mathbf{0}$ as $t \to \infty$. The following theorem shows that the asymptotic stability is achieved when the acceleration-based robust position controller is applied to the regulation control problem of robot manipulators.

***Theorem 3:*** If the final value of the reference trajectory is an equilibrium point and the acceleration-based robust position controller is designed by using ***Theorem 2***, then the asymptotic stability is achieved.

*Proof:* ***Theorem 2*** shows that the stability of the acceleration-based robust position control system is achieved if Eq. (19) is satisfied. If the reference of the equilibrium point is generated by using $\ddot{\mathbf{q}}^{ref}, \dot{\mathbf{q}}^{ref} \to \mathbf{0}$, $\mathbf{q}^{ref} \to \mathbf{q}^{fixed}$ and $\dot{\mathbf{q}}^{fixed} = \mathbf{0}$, then the stable position control system should reach at a constant point as $t \to \infty$.

If ***Theorem 2*** holds and the robot reaches at a constant point as $t \to \infty$, then $\mathbf{e_D}$ becomes a constant vector and

$$\mathbf{e_D}^T \boldsymbol{\psi} \geq g_{DOb} \mathbf{e_D}^T \mathbf{M_n} \mathbf{e_D} = \phi \qquad (25)$$

where $\phi$ is a positive constant value. ***Theorem 1*** shows that the origin is asymptotically stable, since the mapping $-\mathbf{e_D} \to -\boldsymbol{\psi}$ is passive. Q.E.D.

## V. SIMULATIONS AND EXPERIMENTS

In this section, the validity of the proposal is verified via simulations and experiments. In simulations, the dynamic model of a 6R robot manipulator was used [25]. In experiments, a revolute joint 3-DOF planar redundant robot manipulator was used (See Fig. 3). The parameters of the planar robot manipulator are given in Table I. The robot was controlled via a real-time Linux OS with the sampling rate of 1 kHz. It was actuated via Maxon RE40 brushed DC motors with μsc-4A

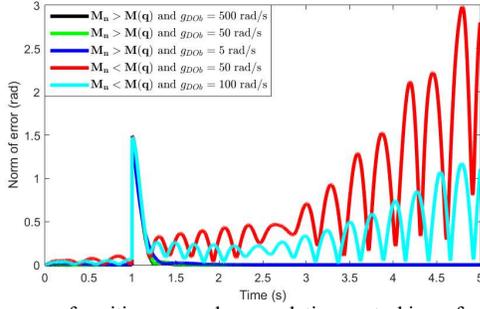

(a) The norm of position error when regulation control is performed.

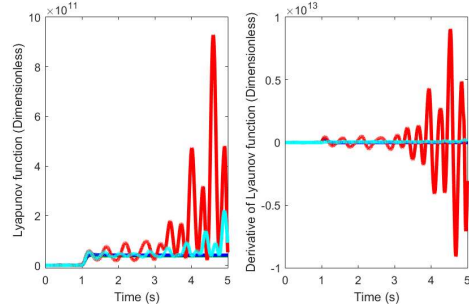

(b) Lyapunov function and its derivative when regulation control is performed. The legends of this figure are given in (a).

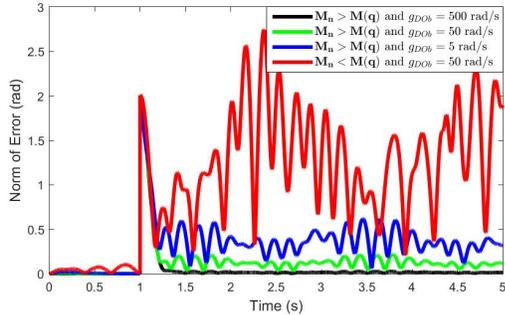

(c) The norm of position error when trajectory tracking control is performed.

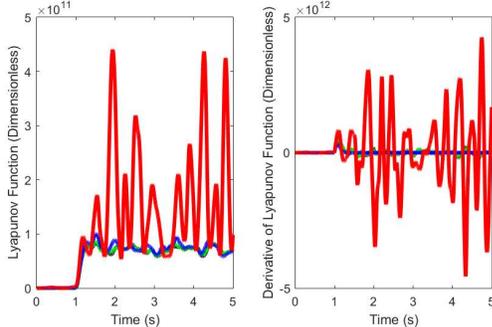

(d) Lyapunov function and its derivative when trajectory tracking control is performed. The legends of this figure are given in (c).

Fig. 4. Regulation and trajectory tracking control results of a 6R robot manipulator. $K_D=25$ and $K_P=250$.

Vanguard systems motor drivers. Data acquisition was carried out with Interface 3340 and 6205C digital-to-analog converter and counter boards. The resolutions of the incremental encoders were 1024 PPR at the first and second joints and 5400 PPR at the last joint. The gear ratios of the speed reducers were 1/26, 1/26 and 1/3 in the first, second and third joints, respectively. The velocity and position control gains of the performance controller were $K_D=80$ and $K_P=1600$, respectively. Velocity was estimated by using the pseudo-derivative of position measurement, and its bandwidth was 1000 rad/s.

### A. Simulations

Fig. 4a and Fig. 4b illustrate the regulation control results of the acceleration-based robust position control system. Step reference inputs, whose magnitudes are 25cm and 40cm, were applied in the x and y operational space coordinates, respectively. The figures show that the norm of the error diverges, i.e., the system becomes unstable, when the robust position controller is designed by using a small nominal inertia matrix and the bandwidth of DOb cannot compensate the disturbances due to inertia variation. The stability can be simply improved by increasing either the bandwidth of DOb or the nominal inertia matrix. If the robust position controller is designed by using $\mathbf{M_n} > \mathbf{M(q)}$, then the asymptotic stability can be achieved for the different bandwidth values of DOb. It is clear from the figure that not only the robustness against disturbances but also the stability is improved by increasing the bandwidth of DOb. The Lyapunov function converges to a constant value if the asymptotic stability is achieved.

Fig. 4c and Fig. 4d illustrate the trajectory tracking control results of the acceleration-based robust position control system. A circular reference trajectory, whose radius is 20 cm, was applied in the x and y operational space coordinate frames. Similarly, the stability is improved by increasing the bandwidth of DOb and/or the nominal inertia matrix. However, the asymptotic stability cannot be achieved in trajectory tracking control. The norm of the position control error is bounded if the stability is achieved and shrinks as the bandwidth of DOb and/or the nominal inertia matrix are increased.

### B. Experiments

In this subsection, the nominal inertia matrix is defined by using

$$\mathbf{M_n} = \mathbf{M_n^d} + \mathbf{M_n^{nd}} = \begin{bmatrix} 0.0332 & 0.00181 & 0.00573 \\ 0.00181 & 0.0163 & 0.00367 \\ 0.00573 & 0.00367 & 0.00117 \end{bmatrix} \quad (26)$$

where $\mathbf{M_n^d}$ and $\mathbf{M_n^{nd}}$ represent the diagonal and non-diagonal inertia matrices, respectively. The bandwidth of DOb is 200 rad/s if it is not specified.

Fig. 5a and Fig. 5b illustrate the acceleration-based robust position control experimental results of the planar robot manipulator when $\mathbf{g(q)}=\mathbf{0}$ and $\mathbf{g(q)} \neq \mathbf{0}$, respectively. An unknown external disturbance was applied by directly pushing the robot's tip point where a force sensor was mounted to measure the applied disturbance. A diagonal nominal inertia matrix was used in the design of the robust motion controller. The bandwidth of DOb and the nominal inertia matrix were experimentally tuned to adjust the robustness and noise sensitivity. As they are increased, the robust stability and performance of the position control system are improved yet the noise sensitivity deteriorates; i.e, there is a trade-off between the robustness and noise sensitivity in the proposed position controller. It is clear from the figures that the planar robot can precisely track its reference trajectories in operational space by cancelling/ suppressing internal and external disturbances when the acceleration-based robust position controller is implemented. Fig. 5c and Fig. 5d respectively illustrate the force measurement and the estimation of disturbances when $\mathbf{g(q)}=\mathbf{0}$ and $\mathbf{g(q)} \neq \mathbf{0}$ in regulation control. The difference

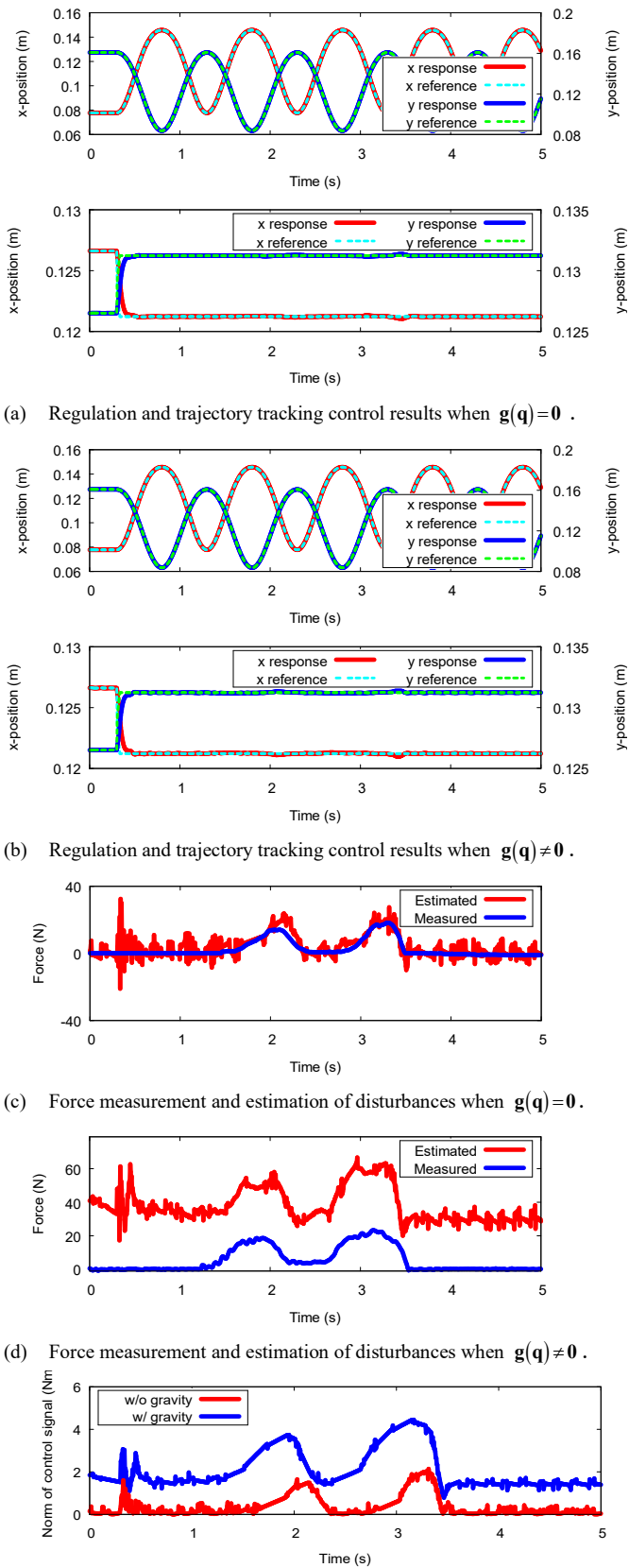

(a) Regulation and trajectory tracking control results when $\mathbf{g(q)}=\mathbf{0}$.

(b) Regulation and trajectory tracking control results when $\mathbf{g(q)}\neq\mathbf{0}$.

(c) Force measurement and estimation of disturbances when $\mathbf{g(q)}=\mathbf{0}$.

(d) Force measurement and estimation of disturbances when $\mathbf{g(q)}\neq\mathbf{0}$.

(e) Control signals.

Fig. 5. Robust regulation and trajectory tracking control experiments when $\mathbf{g(q)}=\mathbf{0}$ and $\mathbf{g(q)}\neq\mathbf{0}$.

between the force measurement and disturbance estimation is due to the internal disturbances that are estimated by DOb. In other words, a DOb can be used as a force sensor (i.e., external disturbance estimator) if internal disturbances are well-suppressed. Fig. 5e illustrates the control signal of the robust position controller when regulation control is performed.

Let us now focus on the means of tuning inertia matrix and its effects on the robust stability and performance of the ABC system. Fig. 6 illustrates the robust regulation control results when different diagonal nominal inertia matrices are used in the design of the position controller. A step reference input was applied at 0.25 s, and a step external disturbance was applied at 1 s. Fig. 6a and Fig. 6b show that not only the stability but also the robustness against disturbances deteriorates as the nominal inertia matrix is decreased. Since the bandwidth of DOb is high enough to compensate inertia variations, the asymptotic stability is achieved in steady state. However, the transient response significantly deteriorates by the poor stability and robustness. Fig. 6c shows that as the nominal inertia matrix is increased, the noise sensitivity of the position control system deteriorates. Therefore, not only the bandwidth of DOb but also the nominal inertia matrix should be properly tuned by considering the robustness, stability and noise sensitivity of the acceleration-based position control system. Fig. 6d illustrates the control signal of the robust position controller.

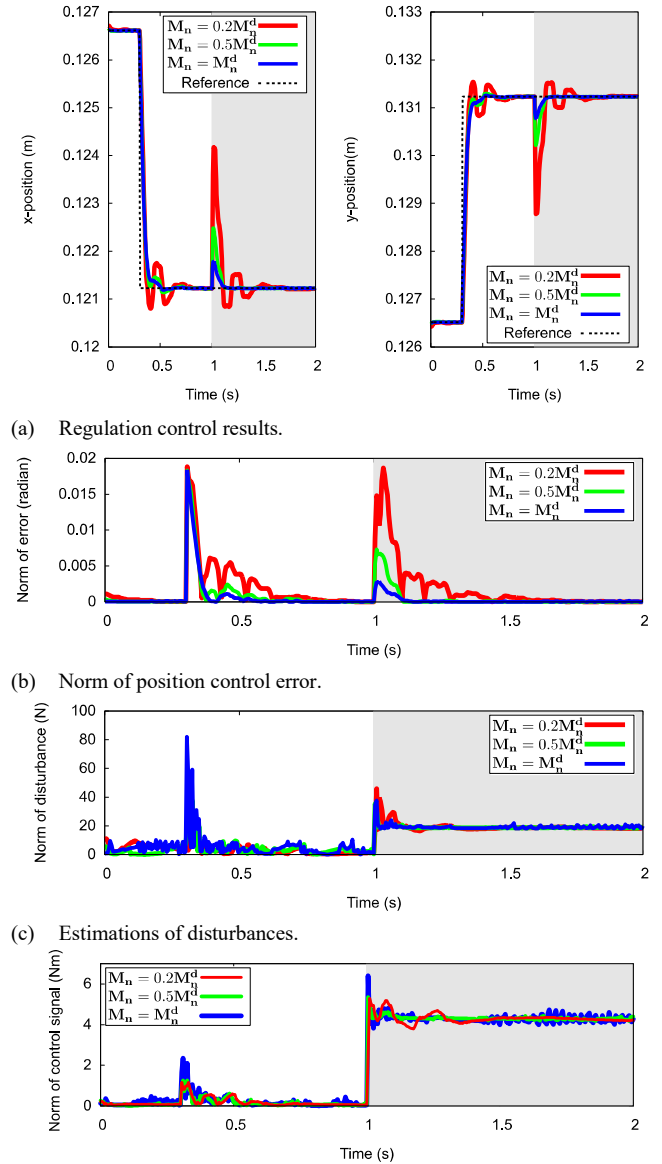

(a) Regulation control results.

(b) Norm of position control error.

(c) Estimations of disturbances.

(d) Control signals.

Fig. 6. Robust regulation control experiments when different diagonal nominal inertia matrices are used in the design of DOb.

Fig. 7 shows the regulation control results when DOb is designed by using not only the diagonal but also the non-diagonal elements of the nominal inertia matrix. Similarly, a step reference input was applied at 0.25s, and a step external disturbance was applied at 1 s. As it is proved in **Theorem 2**, Fig. 7 shows that the stability and robustness against disturbances are improved as the nominal inertia matrix is increased; however, the controller becomes more noise sensitive.

Fig. 8 illustrates the robust regulation control results when the same reference input and external disturbance are applied. In this experiment, different nominal inertia matrices were used in the design of DOb. As the non-diagonal terms of the nominal inertia matrix were increased, the robustness was improved without degrading the noise sensitivity. Using the non-diagonal elements of the nominal inertia matrix provides us extra control parameters to adjust the trade-off between the robustness and noise sensitivity of the ABC system.

Last, let us consider the influence of the bandwidth of DOb and nominal inertia matrix on the stability of the acceleration-based robust position control system. Fig. 9 shows the robust trajectory tracking control results when different diagonal nominal inertia matrices and bandwidth values are used in the design of DOb. It is clear from the figure that the stability can be improved by increasing either the bandwidth of DOb or the nominal inertia matrix. As the bandwidth of DOb is increased, the inertia variations can be compensated and the stability can be achieved even DOb is designed by using $\mathbf{M_n} < \mathbf{M}(\mathbf{q})$.

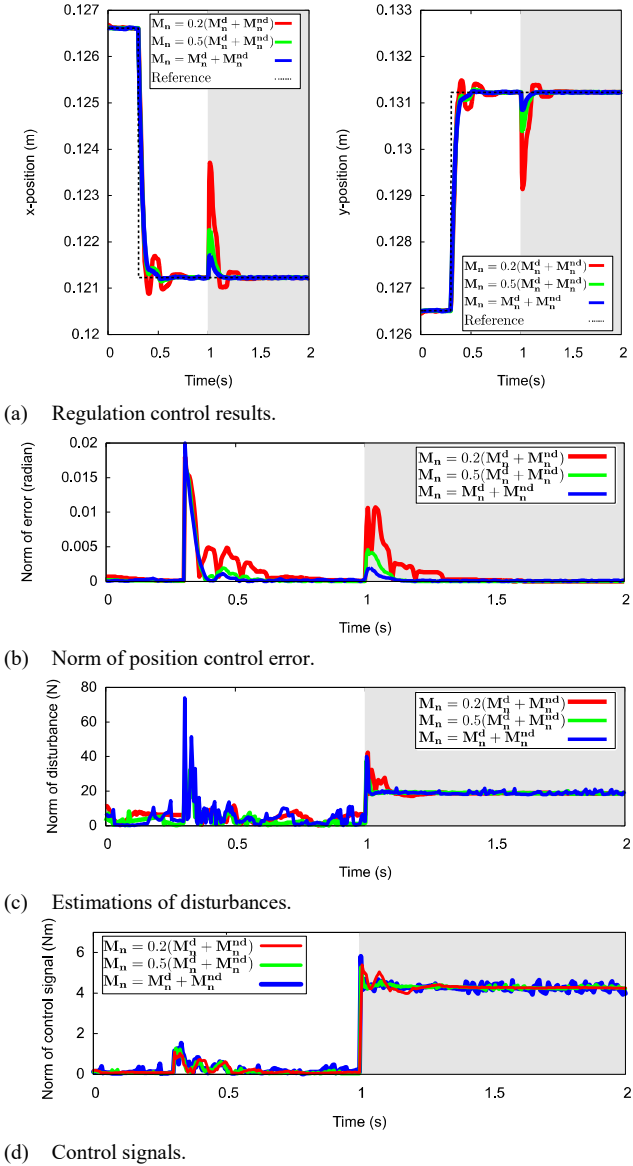

(a) Regulation control results.

(b) Norm of position control error.

(c) Estimations of disturbances.

(d) Control signals.

Fig. 7. Robust regulation control experiments when different nominal inertia matrices are used in the design of DOb.

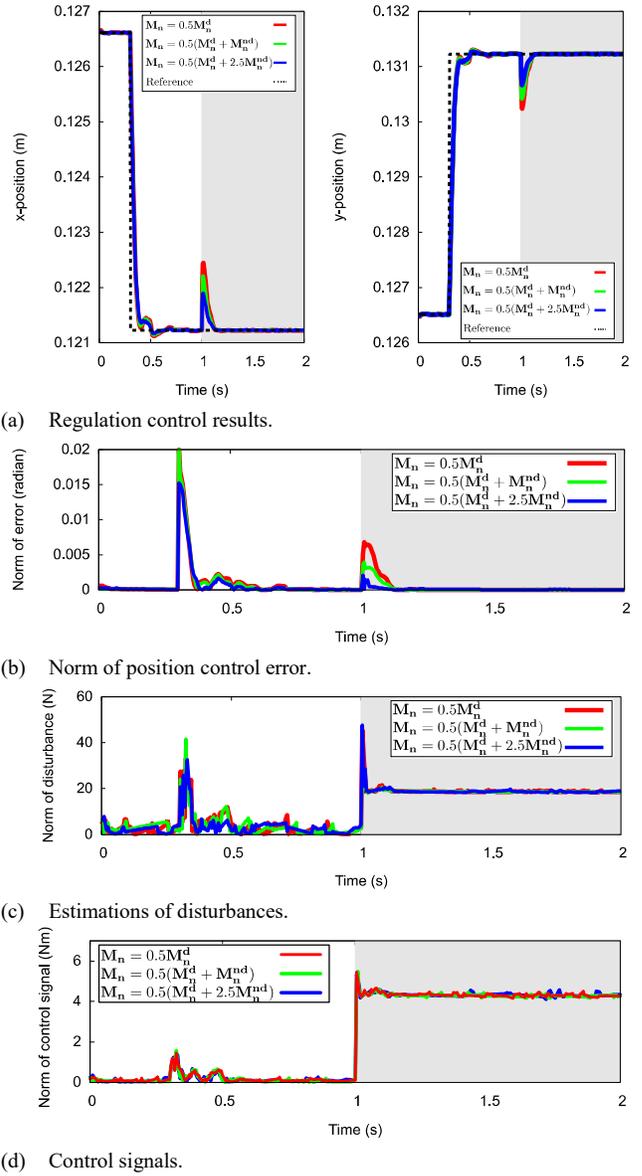

(a) Regulation control results.

(b) Norm of position control error.

(c) Estimations of disturbances.

(d) Control signals.

Fig. 8. Robust regulation control experiments when different nominal inertia matrices are used in the design of DOb. Diagonal and non-diagonal inertia matrices are compared.

However, if the bandwidth of DOb is not high enough to compensate inertia variations, then the robust position control system becomes unstable. To improve the stability of the robust position control system by eliminating the influence of inertia variations, adaptive control methods can be applied by online identifying the inertia matrix of the robot manipulator [26]. In the adaptive ABC, the noise-sensitivity may influence the convergence of parameter identification, and thus the stability of the position control system. Robust identification methods can be used to improve the stability of the control system [27].

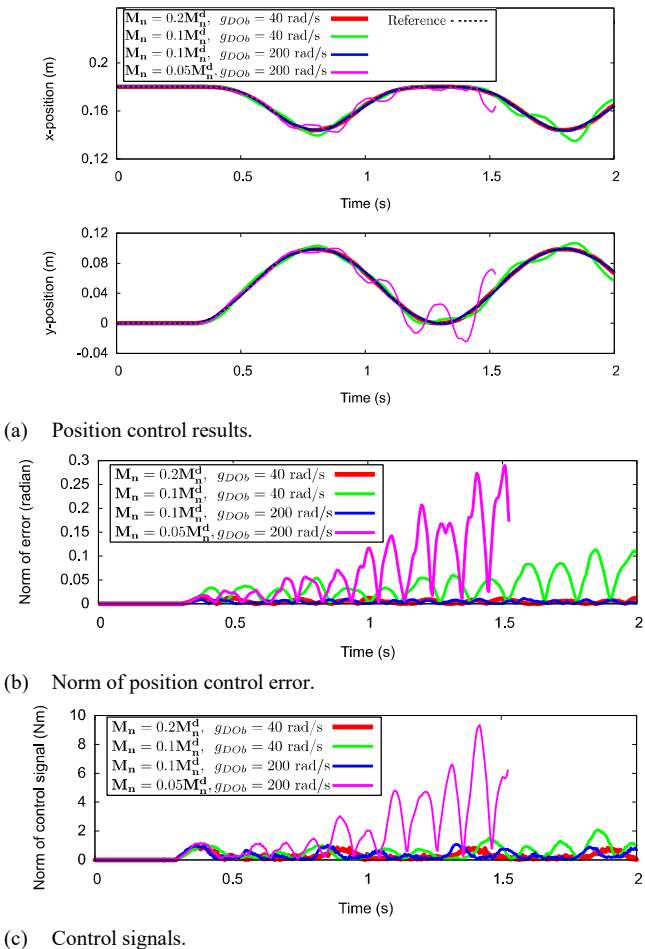

(a) Position control results.

(b) Norm of position control error.

(c) Control signals.

Fig. 9. Robust trajectory tracking control result when different diagonal nominal inertia matrices and bandwidth values are used in the design of DOb.

## VI. CONCLUSION

In this paper, a new nonlinear stability analysis has been proposed for the acceleration-based robust position control of robot manipulators. The proposed controller has two-degrees-of freedom; i.e., the robustness and performance can be independently adjusted by using DOb and PD controller in the inner and outer loops, respectively. If the robust motion controller is properly tuned as shown in the paper, then the position error is uniformly ultimately bounded in trajectory tracking control and asymptotically goes to zero in regulation control. ***Theorem 2*** shows that the position control error can be directly lessened, i.e., the stability and performance can be improved, by increasing either the bandwidth of DOb or the nominal inertia matrix. However, they are limited by practical design constraints, such as noise, as shown in Section V.

It is theoretically and experimentally proved that the diagonal and non-diagonal terms of the nominal inertia matrix can be used to tune the dynamic response of the acceleration-based robust position control system. As it is shown in Fig. 8, the robustness can be improved without degrading the noise sensitivity when the non-diagonal terms of the nominal inertia matrix are properly tuned.

***Theorem*** 2 and ***Theorem*** 3 provide us a clear insight into the robust stability and performance of the acceleration-based robust position control systems. However, the design of the nominal inertia matrix is still not straightforward. An optimal nominal inertia matrix design, which considers the trade-off between the robustness and noise sensitivity, should be further studied to improve the DOb-based robust motion control systems in practice.